\documentstyle[12pt,epsf]{article}
\def\slashchar#1{\setbox0=\hbox{$#1$}           
   \dimen0=\wd0                                 
   \setbox1=\hbox{/} \dimen1=\wd1               
   \ifdim\dimen0>\dimen1                        
      \rlap{\hbox to \dimen0{\hfil/\hfil}}      
      #1                                        
   \else                                        
      \rlap{\hbox to \dimen1{\hfil$#1$\hfil}}   
      /                                         
   \fi}                                         %
\def\mEt{\mbox{${\hbox{$E$\kern-0.6em\lower-.1ex\hbox{/}}}_T$}\, } 

\def\MU{M_{\rm U}}
\def\MP{M_{\rm P}}
\def\MX{M_{\rm X}}

\def\stilde{\widetilde}

\def\conj{{\rm c.c.}}
\newcommand{\newc}{\newcommand}
\newc{\lcal}{\int {\cal L}dt}
\newc{\gsim}{\lower.7ex\hbox{$\;\stackrel{\textstyle>}{\sim}\;$}}
\newc{\lsim}{\lower.7ex\hbox{$\;\stackrel{\textstyle<}{\sim}\;$}}
\def\beq{\begin{eqnarray}}
\def\eeq{\end{eqnarray}}
\def\bea{\begin{eqnarray}}
\def\eea{\end{eqnarray}}
\catcode`@=11
\long\def\@caption#1[#2]#3{\par\addcontentsline{\csname
  ext@#1\endcsname}{#1}{\protect\numberline{\csname
  the#1\endcsname}{\ignorespaces #2}}\begingroup
    \small
    \@parboxrestore
    \@makecaption{\csname fnum@#1\endcsname}{\ignorespaces #3}\par
  \endgroup}
\catcode`@=12

\def\dofigs#1#2#3{\centerline{\epsfxsize=#1\epsfbox{#2}%
   \hfil\epsfxsize=#1\epsfbox{#3}}}

\begin{document}
\begin{titlepage}
\begin{flushright}
hep-ph/0106280 \\
FERMILAB-Pub-01/190-T
\end{flushright}
\vspace{0.3in}

\baselineskip=20pt
\begin{center}

{\large\bf
Predictions of the sign of $\mu$ from supersymmetry breaking models}

\end{center}

\vspace{.15in}
\begin{center}

{\sc Stephen P.~Martin}

\vspace{.1in}
{\it Department of Physics, Northern Illinois University, DeKalb IL 60115
{$\rm and$}\\
}{\it Fermi National Accelerator Laboratory,
P.O. Box 500, Batavia IL 60510 \\} 
\end{center}

\vspace{0.8in}

\begin{abstract}\noindent The sign of the supersymmetric Higgs mass $\mu$
is usually taken as an independent input parameter in analyses of the
supersymmetric standard model. I study the role of theories of
supersymmetry breaking in determining the sign of $\mu$ as an output. 
Models with vanishing soft scalar couplings at the apparent
gauge coupling unification scale are known to predict positive $\mu$.
I investigate more general
results for the sign of $\mu$ as a function of the holomorphic soft scalar
couplings, and compare to predictions of
models with gaugino mass dominance at higher scales. In a significant
region of the
$B_0/m_{1/2}$ versus $A_0/m_{1/2}$ plane
including $A_0 = B_0 = 0$, $\mu$ must be positive. In another
region, $\mu$ is definitely negative. Only in a smaller intermediate
region does knowledge of the supersymmetry breaking mechanism not permit a
definite prediction of the sign of $\mu$. The last region will shrink
considerably as the top quark mass becomes more accurately known.

\end{abstract}
\end{titlepage}
\baselineskip=17.4pt
\setcounter{footnote}{1}
\setcounter{page}{2}
\setcounter{figure}{0}
\setcounter{table}{0}


In the minimal supersymmetric standard model (MSSM)
\cite{HaberKane,primer}, the Higgs mass term $\mu$ is the only coupling
which does not explicitly break supersymmetry that has not already been
directly measured by experiment. Nevertheless, in phenomenological
treatments of supersymmetric models, it is usual to treat $|\mu|$ as an 
output rather than an input parameter, because it can be
fixed in terms of the other parameters from our knowledge of the
electroweak scale. However, this condition alone
does not address the phase of $\mu$, which is left unfixed by the
conditions of electroweak symmetry breaking (EWSB).
The lack of observed CP violation in the electric dipole moments of the
neutron and electron requires that large relative phases in the MSSM
lagrangian must either be absent or aligned to rather particular values.
Barring the latter possibility, it follows that all gaugino masses should
be (at least nearly) relatively real, and that with appropriately chosen
phase conventions $\mu$ is real and the phases of scalar cubic couplings
are equal to their Yukawa coupling counterparts. 

The remaining discrete phase freedom sign$(\mu)$ is therefore usually
regarded as an independent input parameter.  However, if the mechanism of
supersymmetry breaking is known, the phase of
$\mu$ including its sign is often determined purely from the theory and
knowledge of already-measured dimensionless supersymmetry-preserving
couplings.  This has been noted before in the contexts of flipped
$SU(5)\times U(1)$ no-scale supergravity models \cite{flippednoscale} and
in gauge-mediated supersymmetry breaking
models\cite{gmsbexamples}-\cite{Mafi:2001kg}.  More 
generally,
a complete model of supersymmetry breaking should predict boundary
conditions for all soft parameters in terms of supersymmetric parameters.
This implies that, under many (but not all!)  circumstances, the sign of
$\mu$ should properly be regarded as an output prediction rather than an
input assumption. Conversely, an experimental determination of the sign of
$\mu$ will provide a non-trivial test of different models of supersymmetry
breaking. In this paper I will study the ability of flavor-preserving
high-scale theories of supersymmetry breaking to predict the sign of
$\mu$, and consider under what circumstances such a prediction can be made
unambiguously.

In this paper, it is assumed that the gaugino mass parameters
$M_1$, $M_2$, and $M_3$ indeed have
the same phase, so that they can be taken real and positive without
loss of generality.\footnote{This would follow, for example, in GUT models
in which all gaugino masses are unified.} 
To fix conventions explicitly, 
the tree-level neutral Higgs potential is given by
\beq
V \!&=&\!
(|\mu|^2 + m^2_{H_u}) |H_u^0|^2
+ (|\mu|^2 + m^2_{H_d}) |H_d^0|^2
- (b\, H_u^0 H_d^0 + \conj)
\nonumber \\ && + {1\over 8} (g^2 + g^{\prime 2})
( |H_u^0|^2 - |H_d^0|^2 )^2 ,
\label{higgsv}
\eeq
Here $b$ is the holomorphic soft supersymmetry-breaking Higgs squared mass
parameter. (Other common notations in the literature for this
term are $B\mu$ and $m_{12}^2$ and $m_3^2$.)
Without loss of generality, a suitably renormalized $b$ is taken to be
real and positive
at a renormalization group (RG) scale near or below 1 TeV, to fulfill the
condition that at the minimum
of the effective potential,
the Higgs
fields will have real
positive VEVs:
\beq
\langle H_u^0\rangle = v_u;\qquad \langle H_d^0\rangle = v_d;\qquad
v_u^2 + v_d^2 \approx (175\>{\rm GeV})^2;\qquad v_u/v_d \equiv\tan\beta .
\eeq
The tree-level top, bottom and tau masses and Yukawa couplings
$m_t = v_u y_t$, $m_b = v_d y_b$ and $m_\tau = v_d
y_\tau$ are simultaneously real and positive. (Lighter fermion masses 
are neglected, so CKM CP violation is not an issue.)
Neutralino and chargino mass matrices are given by
\beq
{\bf M}_{\stilde N} = \pmatrix{M_1 & 0 & -g'v_d/\sqrt{2}& g'v_u/\sqrt{2}
\cr
0 & M_2 & gv_d/\sqrt{2} & - gv_u /\sqrt{2} \cr
-g'v_d/\sqrt{2}&gv_d/\sqrt{2} & 0 & -\mu \cr
g'v_u/\sqrt{2} & -gv_u/\sqrt{2} & -\mu & 0 },
\qquad
{\bf M}_{\stilde C} = \pmatrix{M_2 & g v_u\cr
                         g v_d & \mu} .
\eeq
The relevant soft supersymmetry-breaking terms include 
\beq
-{\cal L}_{\rm soft} &=& 
-b H_u^0 H_d^0 + a_t \stilde t_L \stilde t_R^* H_u^0  +
a_b \stilde b_L \stilde b^*_R H_d^0 + 
a_\tau \stilde \tau_L \stilde \tau_R^* H_d^0 
+ {\rm c.c.} ,
\eeq
so that the stop and sbottom squared mass matrices are:
\beq
{\bf m}^2_{\stilde t} &=& \pmatrix{ m^2_{\stilde t_L} + y_t^2 v_u^2 +
D_{\stilde t_L}
& a_t v_u - \mu y_t v_d \cr
a_t v_u - \mu y_t v_d & m^2_{\stilde t_R} + y_t^2 v_u^2 + 
D_{\stilde t_R}};\\
{\bf m}^2_{\stilde b} &=& \pmatrix{ m^2_{\stilde b_L} + y_b^2 v_d^2 +
D_{\stilde b_L}
& a_b v_d - \mu y_b v_u \cr
a_b v_d - \mu y_b v_u & m^2_{\stilde b_R} +  y_b^2 v_d^2 + D_{\stilde
b_R} },
\eeq
where $D_\phi = 
(g^2 T^\phi_3 - g^{\prime 2} Y^\phi)(v_d^2 - v_u^2)/2$. These phase
conventions agree with those in \cite{HaberKane,primer}.

Within the framework of
supersymmetry breaking communicated by arbitrary Planck-suppressed
operators, the assumption that $\mu$ is real is a strong and seemingly
unnatural one, requiring justification in terms of some
organizing principle. 
One way of addressing this is to require
that gaugino
masses are the dominant source of all supersymmetry breaking at some
RG input scale $\MX$.
Other
soft supersymmetry-breaking parameters can then be thought of as radiative
effects due to large logarithms which can be resummed using the
renormalization group. 
Older versions of this idea followed from the
ideas of ``no-scale" supergravity models \cite{noscale,flippednoscale},
and it has
found a
different
justification recently in terms of models with supersymmetry breaking
displaced along compactified extra dimensions 
\cite{Kaplan:2000ac}-%
\cite{Cheng:2001an}.  
A crucial benefit of these models is
that they naturally avoid the
most dangerous types of supersymmetric flavor violation, since the
gaugino interactions which communicate supersymmetry breaking to the
sfermion masses are
automatically flavor-blind.

If gaugino
masses have a
common phase and are the dominant source of supersymmetry breaking, then
it is well-known that $\mu$ can be taken to be real without loss of
generality. One way to understand this is to consider the form of
the RG equations for the holomorphic scalar supersymmetry-breaking
interactions $b$, $a_t$, $a_b$, and $a_\tau$. At all orders in
perturbation theory, these can be written
in the form\cite{allordersbetas}:
\beq
{d\over dt}(a_f/y_f) &=& -2 {\cal O} [\beta(y_f)/ y_f] ,
\label{aallorders}\\
{d\over dt}(b/\mu) &=& -2 {\cal O} [\beta(\mu) / \mu]
\label{ballorders}
\eeq
where
\beq
{\cal O} &\equiv & {1\over 2} \sum_a M_a g_a {\partial \over \partial g_a}
-
\sum_f a_f {\partial \over \partial y_f}
\label{Ooperator}
\eeq
is a differential operator on the space of gauge and holomorphic 
couplings.
The index $a$
labels
the gauge groups with gauge couplings $g_a$ and gaugino
masses $M_a$, and
$t = {\rm ln}(Q/Q_0)$ with $Q$ the RG scale.
If $b/\mu$, $a_t/y_t$, $a_b/y_b$, and $a_\tau/y_\tau$ are
negligible
at the input scale and are generated by radiative corrections, they will
be real at all other scales, since $\cal O$ is linear in
$M_a$ and $a_f$ and the quantities $\beta(y_f)/ y_f$ and $\beta(\mu) /
\mu$ are sums of real superfield anomalous dimensions.
Since $b$, $y_t$, $y_b$, $y_\tau$, and one $M_a$ are real by
convention, and the other $M_a$ are real by assumption, it follows
that $\mu$, $a_t$, $a_b$, $a_\tau$ are real within the same set of
conventions. 

The fact that the running  gauge couplings of the MSSM are found to nearly
meet at a scale near $2 \times 10^{16}$ GeV is suggestive that a
perturbative RG
analysis can be applied for all couplings and parameters up to that scale.
However, whether in models of
extra dimensions, or ``no-scale" models, or supergravity-inspired models
which happen to have gaugino mass domination, it
is likely that the true input scale is higher, perhaps
at the reduced Planck scale $\MP = 2.4 \times 10^{18}$ GeV. It is
difficult to say with any confidence what the RG running should
be like above $\MU$, except that the evolution of soft parameters is
significant and dominated by gaugino mass effects. 
Therefore, it is useful to work with boundary conditions for the gaugino
masses $M_1$, $M_2$, $M_3$ and soft scalar interactions:
\beq
&&b/\mu \equiv B_0 \label{bcB} \\
&&a_t/y_t = A_{0t};\qquad 
a_b/y_b = A_{0b};\qquad
a_\tau/y_\tau = A_{0\tau} \label{bcA}
\eeq
imposed at $\MU \equiv 2 \times 10^{16}$ GeV
(except as noted below).
If gaugino mass domination is input at $\MU$, then one
would have $A_{0t} = A_{0b} = A_{0\tau} = B_0 =0$
at that scale. However, if the true input scale is
higher, then an examination of the perturbative form
of the beta functions eqs.~(\ref{aallorders})-(\ref{Ooperator})
shows that $B_0$ and $A_{0t}$, $A_{0b}$, $A_{0\tau}$ will each be 
negative at
$\MU$ due to loops involving gauginos. 

In general one expects that $A_{0t}$, $A_{0b}$, $A_{0\tau}$ obtain
different corrections from physics above $\MU$, depending on how the MSSM
superfields fit into whatever gauge group may reign in that regime.
Similarly, the non-holomorphic scalar squared masses will not be universal
at
$\MU$ if they occupy different representations of the gauge group. In a
study of the sparticle spectrum, it would be crucial to assume knowledge
of these particulars. However, the results below regarding the sign of 
$\mu$
depend only weakly on the effects of non-universal non-holomorphic scalar
masses, which do not enter directly in the RG equations that can affect
the running of the crucial quantity $b/\mu$. Also, the dependence of the
running of $b/\mu$ on scalar cubic couplings below $\MU$ is mostly due (at
least at small or moderate $\tan\beta$) to the single quantity $A_{0t}$,
which in many models is not very different from $A_{0b}$ anyway. Results
for the case that the gaugino masses do not unify at $\MU$ are beyond the
scope of this paper, but I expect them to behave in a similar way to the
results below as long as the ratios among $M_1$, $M_2$ and $M_2$ are
moderate. Therefore, for concreteness and simplicity I will use the
traditional boundary conditions
\beq
m_{1/2} &=& M_1 = M_2 = M_3;\\
A_0 &=& A_{0t} = A_{0b} = A_{0\tau};\\
m_0^2 &=& m_\phi^2 \qquad({\rm for}\>{\rm all}\>{\rm \phi})
\label{bcm20}
\eeq
as a convenient parameterization of our ignorance regarding the true
boundary conditions at $\MU$. Each model is then characterized by an
overall gaugino mass scale $m_{1/2}$ and ratios $B_0/m_{1/2}$,
$A_0/m_{1/2}$, and $m^2_0/m^2_{1/2}$. In gaugino mass dominated models,
one generally expects the effective $B_0/m_{1/2}$, $A_0/m_{1/2}$ at $\MU$
to be negative and not too large in magnitude.

In practice, the relation between the sign of $\mu$ and the high-scale
boundary conditions is 
accomplished by choosing $\mu$ and $b$ near the electroweak scale
to produce correct EWSB, running them up to $\MU$, and then
iterating to the desired boundary conditions.
I use 2-loop RG equations \cite{2loopRGEs,Jack:1994rk}
for all MSSM parameters. The
conversion of Standard Model $\overline{{\rm MS}}$ quantities to MSSM
$\overline{{\rm DR}}'$ \cite{DRbar,Jack:1994rk} ones, and the 
relation
between pole masses and
running parameters is accomplished using ref.~\cite{Pierce:1997zz}. 
The conditions for EWSB, the values of $v_u$ and $v_d$, and the 
physical masses of
Higgs scalar bosons are calculated using the full one-loop self-energy
corrections plus the leading two-loop effective potential corrections,
namely those proportional to $g_3^2$ \cite{Zhang:1999bm} 
and those quartic in $y_t$ and $y_b$ \cite{Espinosa:2000df}.
The effective potential minimization is performed at an RG scale equal
to
the geometric mean of the stop masses. 
In this paper, values of
$\tan\beta$ are always quoted as the ratio of
running VEVs at $M_Z$ in the non-decoupling $\overline{\rm DR}'$ scheme
in Landau gauge, determined by running according to the one-loop RG
equations\footnote{Note that the quantities on the
right-hand sides of these equations are the negative of the anomalous
dimensions of the
Higgs fields
in the component field formalism (in which auxiliary fields have been
integrated out) and in Landau gauge, and are not equal to the
superfield anomalous dimensions.}\cite{Castano:1994ri}
\beq
{d\over dt}{\rm ln}(v_u)  &=&  {1\over 16\pi^2} \left [-3 y_t^2 + {3\over
4}
g_2^2 + {3\over 20}
g_1^2 \right ];\\
{d\over dt}{\rm ln}(v_d)  &=& {1\over 16\pi^2} \left [ -3 y_b^2 - y_\tau^2
+
{3\over 4} g_2^2 +
{3\over 20} 
g_1^2 \right ]
\eeq
from the scale at which the effective potential is minimized.
The largest uncertainties in the following come from not knowing
the precise values of the top and bottom quark masses and the QCD
coupling. I take central values and allowed ranges as follows:
\bea
\alpha_3^{\overline{\rm MS}}(M_Z) &=& 0.118\pm 0.003\label{myalpha3}; \\
m_b^{\overline{\rm MS}}(M_Z) &=& 2.88 + 16 (0.118 - \alpha_3)
\pm 0.10 \>{\rm GeV}; \label{mymb}\\
m_t^{\rm pole} &=& 174.3 \pm 8.0 \>{\rm GeV}.
\label{mymtop}
\eea
Here $\alpha_3^{\overline{\rm MS}}$ and $m_b^{\overline{\rm MS}}$
are running parameters in the Standard Model with 5 quark flavors. 
The range in the top quark mass is larger than that quoted in
\cite{Groom:2000in}, because of the theoretical uncertainty in relating
the top-quark Yukawa coupling to the pole mass in supersymmetry.

\begin{figure}[t] 
\dofigs{3.5in}{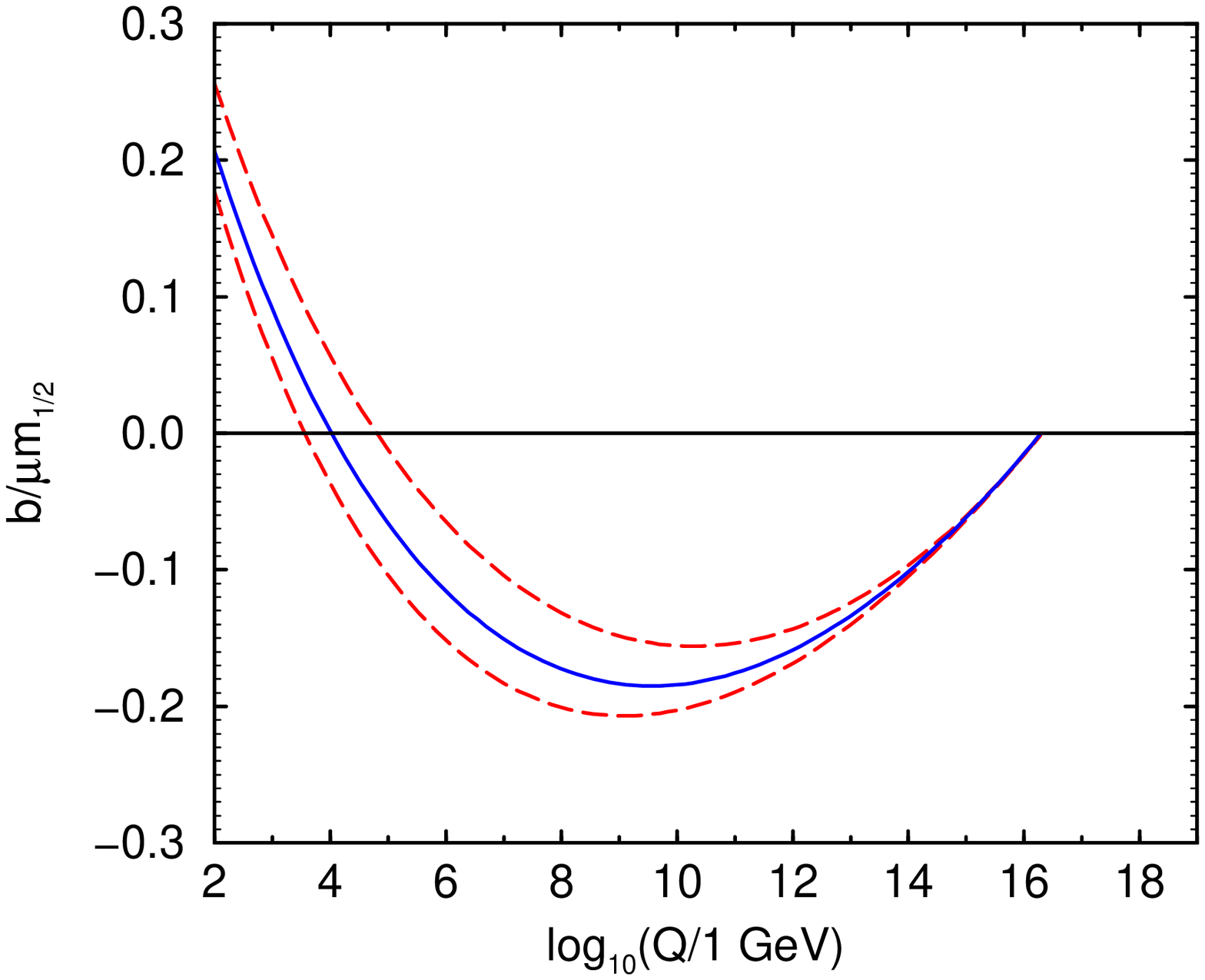}{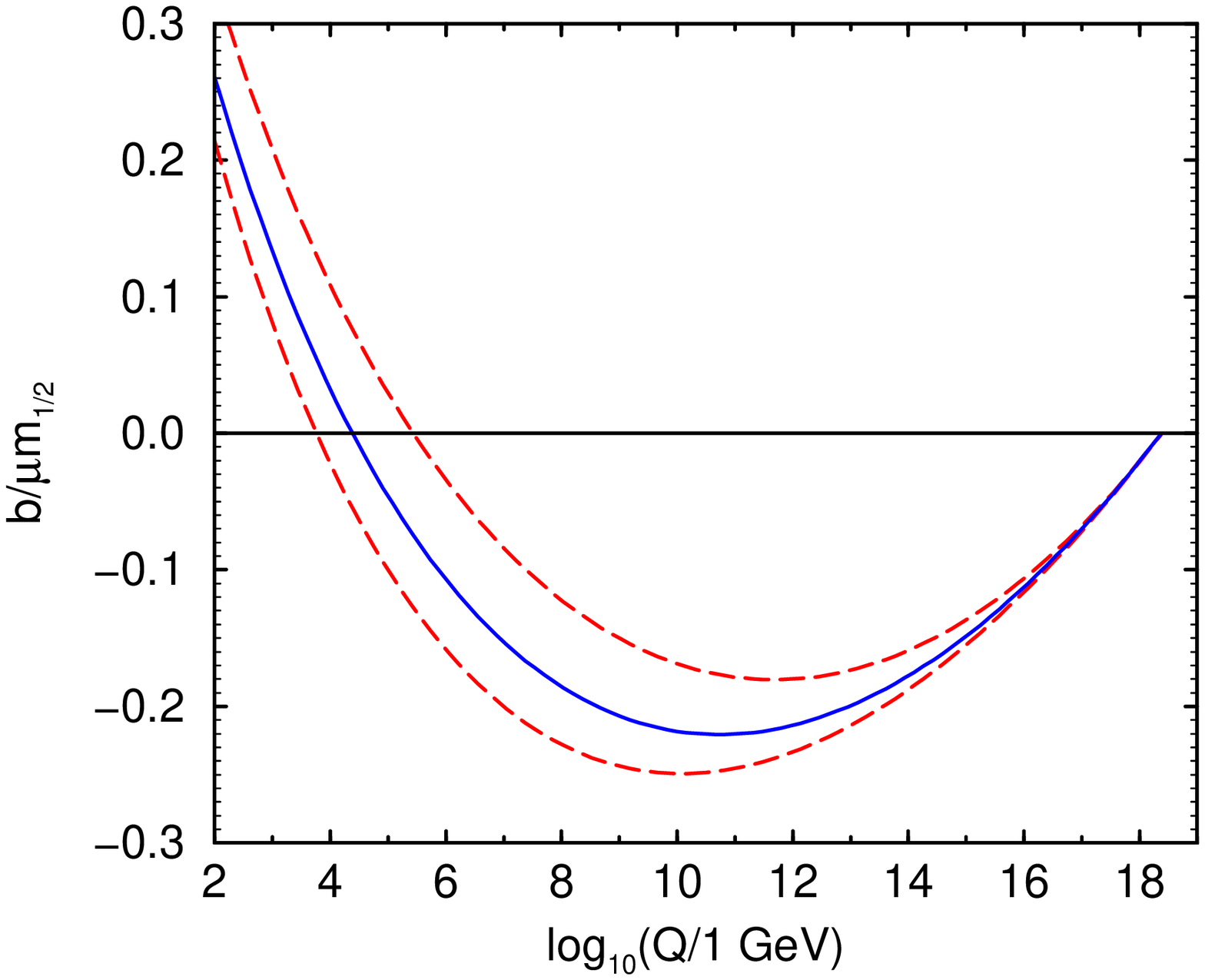}
\caption{Running of the dimensionless ratio of parameters $b/\mu
m_{1/2}$ with the boundary conditions $A_0 = B_0 = m_0 = 0$
and $m_{1/2} = 400$ GeV 
imposed at (a) $\MU = 2 \times 10^{16}$ GeV and (b) $\MP = 2.4 \times
10^{18}$ GeV. 
The solid (blue) lines are
obtained for the central values,
and the dashed (red) lines for the maximum deviation,
implied by eqs.~(\ref{myalpha3})-(\ref{mymtop}). Since $b/\mu m_{1/2}$ is 
positive at the weak scale, $\mu$ must be
positive.}
\label{fig:running}
\end{figure}

The RG evolution of the dimensionless ratio $b/\mu m_{1/2}$ is given in
fig.~\ref{fig:running}(a) for an example
gaugino-mass-dominated model with $A_0 = B_0 = 0$ at $\MX = \MU$.
(The graphs shown also use $m_{1/2}
= 400$ GeV, and $m^2_0 = 0$, but they depend only weakly on those
choices.) With these boundary conditions, $\tan\beta$ is uniquely
determined by the requirements of correct electroweak symmetry breaking,
so there is only one possible RG trajectory for the parameters of the
model once $\alpha_3$, $m_b$ and $m_t$ are fixed.. As
shown, $b/\mu m_{1/2}$ is negative
along most of its evolution towards the infrared, but turns positive at a
scale about two or three orders of magnitude above the electroweak scale.
This can
be
explained as follows. The one-loop RG equations for the holomorphic
soft couplings following from eq.~(\ref{aallorders})-(\ref{ballorders}) 
are:
\beq
16 \pi^2 {d\over dt}(a_t/y_t) &=& 
{32\over 3} g_3^2 M_3 +
6 g_2^2 M_2 + {26 \over 15} g_1^2 M_1 +
12 a_t y_t + 2 a_b y_b;  \label{rgatyt}
\\ 16 \pi^2
{d\over dt}(a_b/y_b) &=& 
{32\over 3} g_3^2 M_3 +
6 g_2^2 M_2 + {14 \over 15} g_1^2 M_1 +
12 a_b y_b + 2 a_t y_t + 2 a_\tau y_\tau;  
\\ 16 \pi^2
{d\over dt}(a_\tau/y_\tau) &=& 
6 g_2^2 M_2 + {18 \over 5} g_1^2 M_1 +
8 a_\tau y_\tau  +
6 a_b y_b ; \label{rgatauytau}
\\ 16 \pi^2
{d\over dt}(b/\mu) &=& 
6 g_2^2 M_2 + {6 \over 5} g_1^2 M_1 +
6 a_t y_t + 
6 a_b y_b +
2 a_\tau y_\tau . \label{rgbmu}
\eeq
At high RG scales, gaugino masses are 
dominant, quickly driving each of $a_f/y_f$ and
$b/\mu$
to negative values in the infrared. Continuing to lower RG scales, the
dominant
contributions to the beta function for $b/\mu$ are the negative ones
proportional to
$a_t y_t$, $a_b y_b$ and $a_\tau y_\tau$. This forces $b/\mu$ 
positive before the electroweak scale is reached. There is a significant
dependence on the top mass and a smaller dependence on the bottom mass
and $\alpha_3$, shown by the envelope of dashed lines. 
Since $b$ is positive near the electroweak scale by convention, the sign
of $\mu$ is the same as the sign of the dimensionless quantity $b/\mu
m_{1/2}$. 
Because there is a unique solution for $\tan\beta$, the
conclusion is that
$\mu$ is inevitably positive.

The model shown in fig.~\ref{fig:running}(a) predicts $\tan\beta$
should be between about 10 (for larger $m_{\rm top}$, corresponding to
the upper dashed line) and 24 (for smaller $m_{\rm top}$, corresponding
to the lower dashed line). It also generally predicts that 
a stau is the lightest supersymmetric particle (LSP), abandoning the
possibility of a supersymmetric source for the cold dark matter. This is
easily corrected if the true input scale is higher than $\MU$. An
example of this is shown in fig.~\ref{fig:running}(b), for which the
scale at which the boundary conditions eq.~(\ref{bcB})-(\ref{bcm20})
with $m_0 = A_0 = B_0=0$
are moved up to the reduced Planck scale $\MP$. For simplicity, no new
particle thresholds are introduced at the apparent unification scale.
As before, the running of $b/\mu$ renders it positive
at the electroweak scale, implying again that $\mu$ must be positive.
In this ultraconservative version of the MSSM with no new particles and
gaugino mass domination at the Planck scale, a bino-like neutralino is the
LSP.

More generally, for given RG trajectories of the dimensionless
supersymmetric parameters, the running of the dimensionless quantity
$b/\mu m_{1/2}$ is determined
uniquely by its boundary condition $B_0/m_{1/2}$ and that of the scalar
cubic couplings, $A_0/m_{1/2}$ at $\MU$. This can be checked from
the form of eqs.~(\ref{aallorders}), (\ref{ballorders}). The effect on the
sign of $\mu$
can be roughly stated as follows. Lowering $A_0/m_{1/2}$ tends to make
the beta function for $b/\mu$ more negative, making $b/\mu
m_{1/2}$ more positive at the weak scale, thus increasing the
parameter space in other variables for which $\mu$ must be positive.
Lowering $B_0/m_{1/2}$ will have the opposite effect, since
for very negative $B_0$, only negative $\mu$ can rescue $b/\mu$ to
make it positive near the electroweak scale. Therefore, one can map out
regions of the $B_0/m_{1/2}$ versus $A_0/m_{1/2}$ which predict that $\mu$
is definitely positive, definitely negative, or can have either sign.

\begin{figure}[t] 
\centering\epsfxsize=5.0in 
\hspace*{0in}
\epsffile{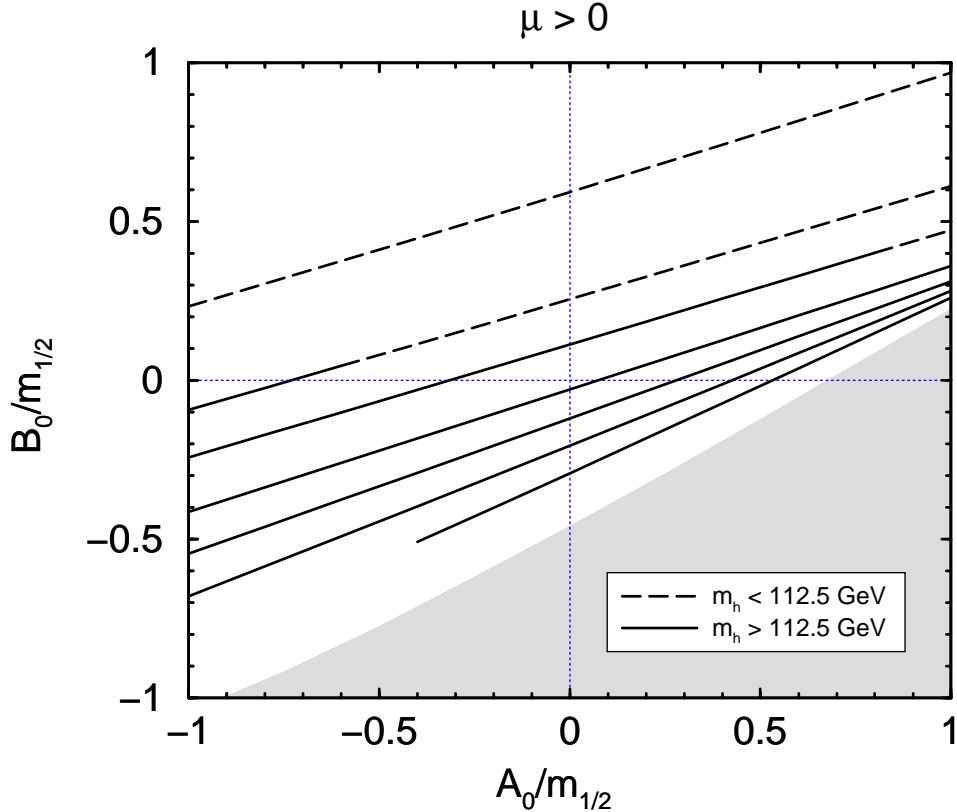}
\caption{The region of the $B_0/m_{1/2}$ vs. $A_0/m_{1/2}$ plane which
allows $\mu > 0$ is unshaded. Boundary conditions are imposed
at $\MU  = 2 \times 10^{16}$ GeV, with
gaugino masses restricted by $m_{1/2} < 400$ GeV and
scalar masses in the range $0< m_0^2/m_{1/2}^2 < 1$. 
All values of
$\tan\beta$ leading to correct EWSB, perturbative couplings up to $M_U$,
and charged superpartners heavier than 100 GeV are
allowed.
Example models lines are shown for
$\tan\beta = 3,6,10,20,30,40,50$
(from top to bottom), with $m_{1/2} = 350$ GeV, $m_0^2/m_{1/2}^2 = 0.5$.}
\label{fig:mupositive}
\end{figure}

Figure \ref{fig:mupositive} shows the region for which $\mu$ can be 
positive, for 
$m_{1/2} < 400$ GeV and $0< m_0^2/m_{1/2}^2 < 1$. In making this
graph, all values of $\tan\beta$ which maintain perturbative couplings up
to $\MU$ are allowed, and the top and bottom quark masses
and $\alpha_3$ are allowed
to
vary over the full ranges indicated in
eq.~(\ref{myalpha3})-(\ref{mymtop}).
All charged sparticle masses are required to be heavier than 100 GeV.
The shaded region indicates where no solution with $\mu > 0$ can be found.
Smaller values of $\tan\beta$ corresponds to points with larger
$B_0/m_{1/2}$, while the largest allowed $\tan\beta$ values occur near the
boundary of the unshaded region. Several example model lines with fixed
$\tan\beta = 3,6,10,20,30,40,50$ are also shown; these were computed with
$m_{1/2} = 350$ GeV, $m_0^2/m_{1/2}^2 = 0.5$
and
central values for
the top and bottom masses and $\alpha_3$. 
I have also indicated by dashed
lines those models for which the lightest CP-even Higgs mass
$m_h$ calculated
as indicated above comes out lighter than 112.5 GeV, for rough comparison
with
LEP2 limits. 
(Even with full one-loop and leading two-loop
calculations, it can be estimated from scale-dependence considerations
that there
is at least a 2 GeV uncertainty in the calculated $m_h$.)
\begin{figure}[t] \centering\epsfxsize=5.0in \hspace*{0in}
\epsffile{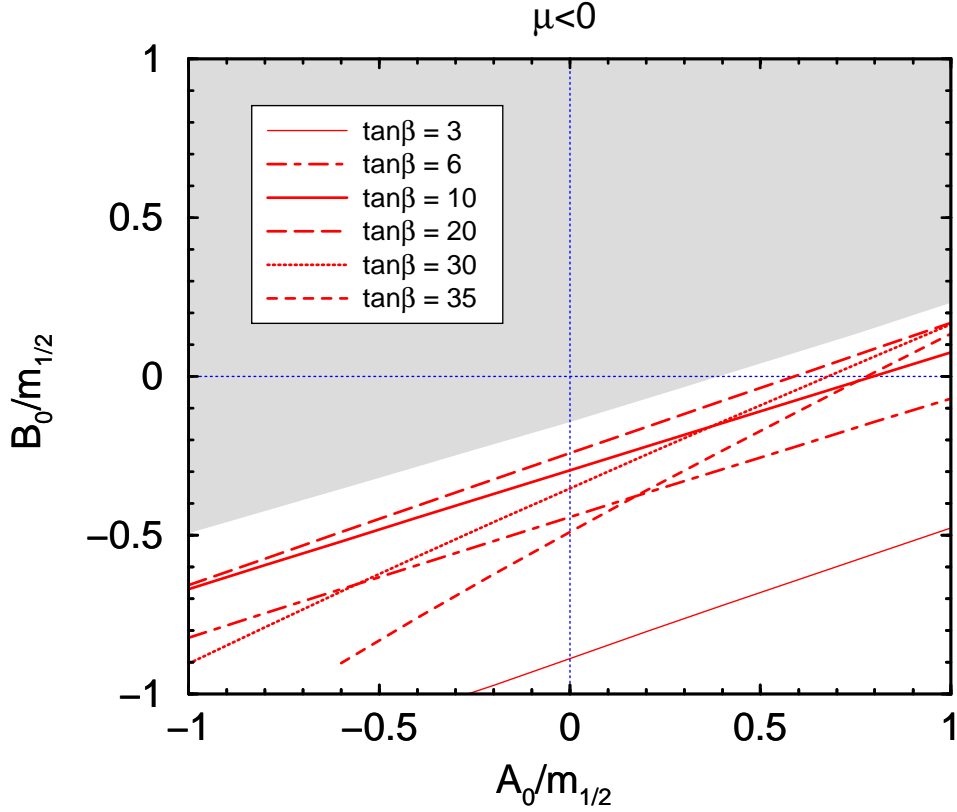} 
\caption{The region of the $B_0/m_{1/2}$ vs. $A_0/m_{1/2}$ plane which
allows $\mu < 0$ is unshaded.  
Universal
gaugino masses are restricted by $m_{1/2} < 400$ GeV and universal scalar
masses lie in the range $0< m_0^2/m_{1/2}^2 < 1$.
All values of
$\tan\beta$ leading to correct EWSB, perturbative couplings up to $M_U$,
and charged superpartners heavier than 100 GeV are
allowed.
Example model lines are shown for
various values of $\tan\beta$, using $m_{1/2} = 350$ GeV, $m_0^2/m_{1/2}^2
= 0.5$.}
\label{fig:munegative} \end{figure}

In contrast, fig.~\ref{fig:munegative} shows the region which allows
negative $\mu$ under the same assumptions.
As suggested by fig.~\ref{fig:running}(a), there is a significant
neighborhood of the point $A_0/m_{1/2} =
B_0/m_{1/2} = 0$ which cannot support negative $\mu$. Here, this is shown 
to be
true for any values of $m_{1/2} <400$ GeV and $0 < m^2_0/m_{1/2}^2 < 1$
and with top and
bottom quark masses and $\alpha_3$ allowed to vary over the entire
ranges indicated in eq.~(\ref{myalpha3})-(\ref{mymtop}).  
Models which approach the border of the allowed region with $\mu<0$
turn out to have intermediate values of $\tan\beta$ (typically between 10
and 25),
while smaller or larger $\tan\beta$ models have larger negative
$B_0/m_{1/2}$.

\begin{figure}[t] 
\centering\epsfxsize=4.2in 
\hspace*{0in}
\epsffile{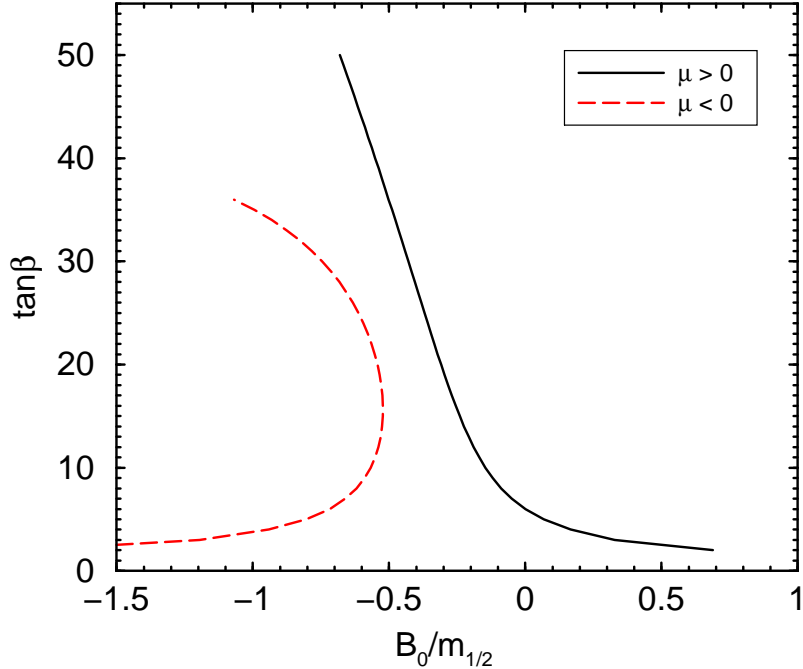}
\caption{Solutions for $\tan\beta$ as a function of $B_0/m_{1/2}$,
for $\mu > 0$ (solid black) and $\mu < 0$ (dashed red).
In each case, the boundary conditions imposed at $\MU = 2\times 10^{16}$
GeV are $A_0 = -0.75 m_{1/2}$, with $m_{1/2} = 400$ GeV and 
$m_0^2/m_{1/2}^2 = 0.5$. This illustrates that for a range of
$B_0/m_{1/2}$
(here, $-0.7$ to $-0.5$), there are sometimes simultaneously distinct 
solutions with
positive $\mu$
and negative $\mu$. For larger $B_0/m_{1/2}$, $\mu$ is definitely positive,
while for smaller $B_0/m_{1/2}$, $\mu$ is definitely negative.}
\label{fig:tanbeta}
\end{figure}

The regions in figs.~\ref{fig:mupositive} and \ref{fig:munegative} allowed
for positive and negative $\mu$ have a significant overlap. 
This represents
a true ambiguity in the sign of $\mu$, even in models for which the
boundary conditions for the soft supersymmetry breaking couplings
are fully specified, and even if the QCD coupling and physical top and
bottom 
masses were known with arbitrary accuracy. To illustrate this,
fig.~\ref{fig:tanbeta} shows solutions for $\tan\beta$ as a function of
a single varying parameter $B_0/m_{1/2}$, with fixed $A_0 = -0.75
m_{1/2}$, $m_{1/2} = 400$ GeV and
$m_0^2/m_{1/2}^2 = 0.5$ and $\alpha_3$, $m_b$ and $m_t$ taking there  
central values.. 
For $B_0/m_{1/2} \gsim -0.5$, there is always only
one solution for $\tan\beta$, corresponding to positive $\mu$. 
For $B_0/m_{1/2} \lsim -0.7$, $\mu$ must be negative, with two distinct
solutions for $\tan\beta$ if $-1.1 < B_0/m_{1/2} <-0.7$. 
For the range $-0.7 < B_0/m_{1/2} < 0.5$, there are
three distinct solutions for $\tan\beta$, one corresponding to positive
$\mu$, and two corresponding to negative $\mu$. 
This is because different sets of Yukawa
couplings $y_t$, $y_b$ and $y_\tau$ can be chosen consistently with
the known masses, with the choice affecting the
running of $b/\mu$. For that range, the sign
of $\mu$ cannot be unambiguously predicted. 

The regions found above can be correlated with particular models
of gaugino mass dominance, depending on the gauge group above $\MU$,
how the MSSM sparticles fit into representations of that group,
and what other particles are present. At one loop order
 in the large-$M_a$ limit,
the RG equations for the soft parameters are
\beq
16 \pi^2
{d\over dt} (b/\mu) &=& 4 \sum_a g_a^2 M_a [C_a(H_u) + C_a(H_d)] ;
\label{rgb}\\
16 \pi^2
{d\over dt} (a_t/y_t) &=& 4\sum_a g_a^2 M_a [C_a(H_u)+C_a(t_L)+C_a(t_R)];
\label{rgat}
\\
16 \pi^2
{d\over dt} (a_b/y_b) &=& 4\sum_a g_a^2 M_a [C_a(H_d)+C_a(b_L)+C_a(b_R)];
\label{rgab}
\\
16 \pi^2
{d\over dt} (a_\tau/y_\tau) &=& 4\sum_a g_a^2 M_a
[C_a(H_d)+C_a(\tau_L)+C_a(\tau_R)];
\label{rgatau}
\\
16 \pi^2
{d\over dt} m_\phi^2 &=& -8 \sum_a g_a^2 |M_a|^2 C_a(\phi) .
\eeq
Here the index $a$ runs over gauge groups with Casimir invariants $C_a$
for the representations of the indicated fields. Now, in principle these
equations could be run down from the input scale to the scale $\MU$ to get
boundary conditions. The resulting one-loop contributions to $b/\mu$ and
$a_f/y_f$ are negative, implying that at the scale $\MU$ we should be in
the lower left quadrant of figs.~\ref{fig:mupositive} and
\ref{fig:munegative}. However, to evaluate these in detail would require a
clairvoyant knowledge of the theory above the apparent unification scale.
Furthermore, in grand unified theory (GUT) models, large representations
generally render perturbation theory invalid below $\MP$. For example, the
minimal missing partner $SU(5)$ model gauge coupling appears to have a
Landau pole if extrapolated at two-loop order, and appears to have an
ultraviolet-stable fixed point at three- and four-loop order
\cite{Martin:2000cr}. The same statement holds for $SO(10)$ models with
large representations. The true UV behavior of such theories is unknown.
Even in models which do not have non-perturbative or Landau-pole behavior
in the gauge couplings, it does {\it not} follow that perturbation theory
for non-holomorphic scalar squared masses is reliable. In fact it is
commonplace for two-loop contributions to non-holomorphic scalar squared
masses to overwhelm the one-loop contributions even if the gauge couplings
remain perturbative.  Another complication is that higher loop corrections
are not linear in quadratic Casimir invariants for $b/\mu$ or $a_f/y_f$.

However, one can still use eqs.~(\ref{rgb}) and (\ref{rgatau}) to
get a rough idea of what to expect for the ratios of $a_f/y_f$
to $b/\mu$ at $\MU$,
at least in the limit of perturbative couplings and small particle
content. 
For example, if
the GUT gauge group is $E_6$ with all MSSM
chiral superfields in ${\bf 27}$ representations, then one finds that
$A_0/B_0 = 3/2$ if one neglects higher loop effects. If the GUT group
is $SO(10)$ with $H_u$ and $H_d$ in a ${\bf 10}$ and top, bottom and tau
in a ${\bf 16}$, then\cite{Barbieri:1995tw,Schmaltz:2000gy}
$A_0/B_0 = 7/4$.
In the case of $SU(5)$ with 
$H_u$ and $H_d$ in ${\bf 5 + \overline 5}$ and standard assignments
for MSSM quarks and leptons, there is a different ``$A_0$"
for top and bottom and tau, 
with\cite{Barbieri:1995tw,Schmaltz:2000gy}
 $A_{0t}/B_0 = 2$ and $A_{0b}/B_0 = A_{0\tau}/B_0 =7/4$. 
The model-dependence tends to cancel out of those ratios
even beyond leading order.
For other non-unified gauge-groups, 
one can make the approximation
that the gauge couplings and gaugino masses above $\MU$ are nearly the
same. For $SU(4) \times SU(2)_L \times U(1)_R$, that would imply $A_0/B_0
= 23/8$. Similarly, with the MSSM gauge group $SU(3)_C\times SU(2)_L \times
U(1)_Y$, with
all gauge couplings and gaugino masses taken as equal above $\MU$, one
would find
$A_{0t}/B_0 = 23/9$ and $A_{0b}/B_0 = 22/9$. This
naive estimate from counting
Casimir invariants
actually agrees reasonably
well with values obtained at $\MU$ for the slightly different
situation
depicted in
fig.~\ref{fig:running}(b), in which all couplings were assumed to run
up to $\MP$ independently according to their MSSM RG equations; there I
found numerically at two loops that $A_{0t}/B_0 \approx A_{0b}/B_0 \approx
2.6$.
Although these ratios
can be modified by many model-dependent effects, one can
take them as suggestive scenarios; respectively, ``$E_6$-like",
``$SO(10)$-like", etc. Summarizing:
\beq
(A_{0t}/B_0,\>A_{0b}/B_0) = 
\left \{ \begin{array}{ll}
(1.5,\>1.5) & \>\>\> E_6{\rm -like}\\
(1.75,\>1.75) & \>\>\>SO(10){\rm -like}\\
(2.0,\>1.75) & \>\>\>SU(5){\rm -like}\\
(2.56,\>2.44) & \>\>\>{\rm MSSM-like}\\
(2.88,\>2.88) & \>\>\>SU(4){\rm -like} .
\end{array}  
\right.
\eeq

\begin{figure}[t] 
\centering\epsfxsize=5.0in 
\hspace*{0in}
\epsffile{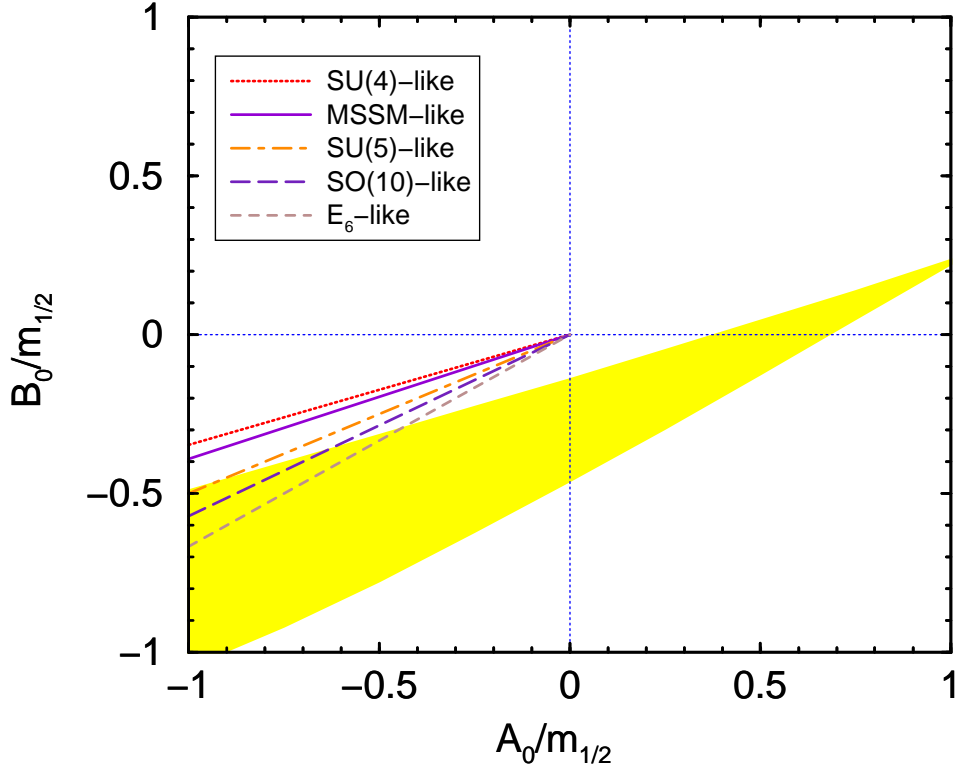}
\caption{The $B_0/m_{1/2}$ vs. $A_0/m_{1/2}$ plane is divided into three
regions, according to whether $\mu$ must be positive (upper unshaded),
$\mu$
must be negative (lower unshaded), or $\mu$ can have either sign
(shaded). The top and bottom quark masses and $\alpha_3$
are allowed to vary over the entire ranges indicated in
eq.~(\ref{myalpha3})-(\ref{mymtop}).  Universal gaugino masses are
restricted by $m_{1/2} < 400$ GeV and universal scalar masses lie in the
range $0< m_0^2/m_{1/2}^2 < 1$. 
All values of $\tan\beta$ leading to
correct EWSB, perturbative couplings up to $M_U$, and
charged superpartners heavier than 100 GeV are allowed.  
For
comparison, the approximate boundary condition ratio predictions of
various model frameworks as described in the text are indicated by lines.}
\label{fig:models}
\end{figure}

These considerations are compared to the preceding general results 
in Figure \ref{fig:models},
which divides the $B_0/m_{1/2}$ vs. $A_0/m_{1/2}$ plane into three
regions. 
In the upper unshaded region including $A_0 = B_0 = 0$,
the sign of $\mu$ is definitely positive. In the lower unshaded region, 
the sign of $\mu$ is
definitely negative. In the intermediate
(yellow) shaded region, the sign of $\mu$ can be either positive
or negative, depending on the values of supersymmetric dimensionless
couplings. The extent of this region was maximized by scanning over
the full allowed range of top and bottom masses and QCD coupling,
as in eqs.~(\ref{myalpha3})-(\ref{mymtop}), as well as including all
$m_{1/2} < 400$ GeV and $0<m_0^2/m_{1/2}^2 < 1$. (The region will grow
very slowly
as the maximum allowed $m_{1/2}$ is increased.)
Also shown are lines corresponding to the boundary conditions
of the different types of models as described above. In the MSSM-like and
$SU(5)$-like cases, $A_{0t}/B_0$ and $A_{0b}/B_0$ and $A_{0\tau}/B_0$ are
slightly different, so the more important factor $A_{0t}/m_{1/2}$ is used.
We learn the following general
lessons. First, 
if the $\MX \rightarrow \MU$ corrections are not too large, then $\mu$
must be positive in all cases. Second, in models with larger
corrections, gauge groups in which the top and bottom quarks
are in larger representations than the Higgs fields require positive $\mu$,
while the highly unified groups $E_6$ and $SO(10)$ can sometimes allow
either sign
of $\mu$. 

\begin{figure}[t] 
\centering\epsfxsize=5.0in 
\hspace*{0in}
\epsffile{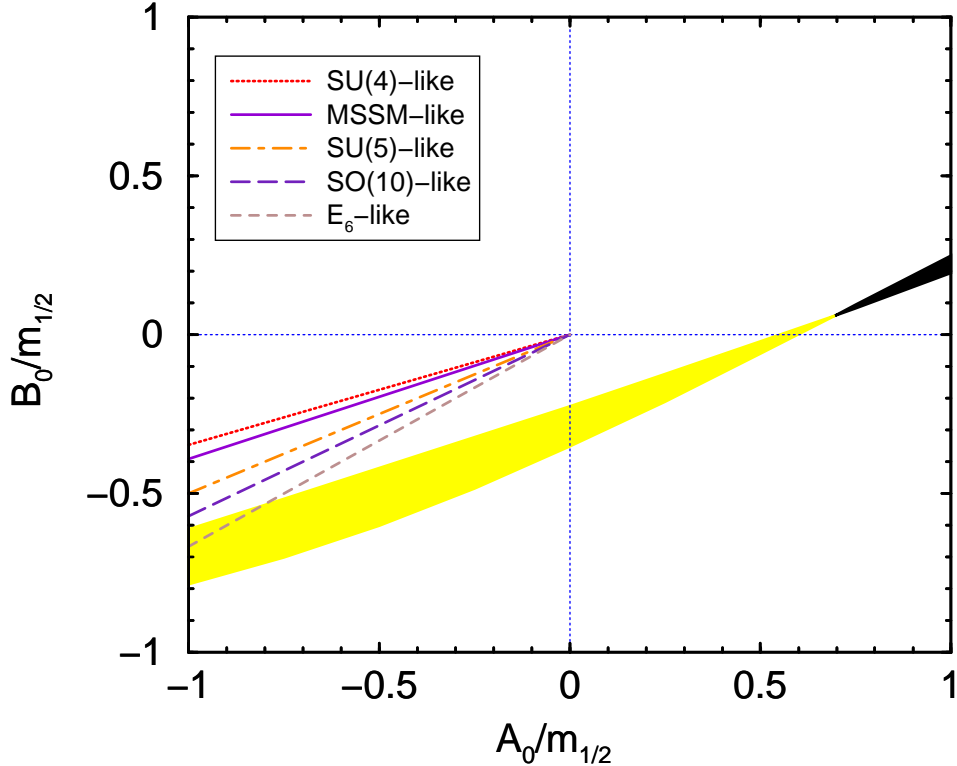}
\caption{As in Figure \ref{fig:models}, but now taking the top quark
mass fixed at its central value, and $m_{1/2} = 400$ GeV with $m_0^2 = 0.5
m_{1/2}^2$. The
upper unshaded region requires $\mu > 0$; the lower unshaded region
requires $\mu<0$; the lighter shaded region allows $\mu$ of either sign;
and the black region at the right allows no solutions. This shows
the improvement that could follow from knowing the top quark mass
accurately and the sfermion masses with reasonable precision.}
\label{fig:models2}
\end{figure}
Someday, the top-quark mass will be better known, and its relation to the
top Yukawa coupling in supersymmetry more accurately computed.
Furthermore, measurements of the sparticle spectrum will enable
determination of $m_{1/2}$, $m_0$. Figure \ref{fig:models2} depicts how
the
situation will improve, now assuming as fixed the present central value
for the top mass with the one-loop supersymmetric corrections, and $m_0^2$
equal to $0.5 m^2_{1/2}$. As shown, the region in which the sign of $\mu$ 
is
not determined by $A_0/m_{1/2}$ and $B_0/m_{1/2}$ shrinks significantly in
this case compared to fig.~\ref{fig:models}. It will shrink even more if
$\tan\beta$ is measured.
This represents a concrete
benefit of an
accurate measurement of the top-quark mass and couplings in testing our
ideas of high-scale physics.

The fact that the $\mu$-term is apparently of the same order of magnitude
as the supersymmetry-breaking soft terms is a major puzzle within
the MSSM. Therefore one should question whether the origin of the
$b$-term might be qualitatively different from that of the other soft
terms, so that the boundary condition $B_0=0$ should not be applied.
However, the origin of the $b$-term cannot be completely arbitrary,
or else one would expect CP-violating couplings in the neutralino and 
chargino
sector. In any case, with a theory for the origin of the $b$-term
one can simply look at the plots above with $B_0/m_{1/2}$ displaced by the
appropriate amount. 

One general strategy for solving the $\mu$ problem relies on replacing it
by the VEV of an additional gauge singlet field \cite{nmssm}. This
allows
all dimensionful parameters to be banned from the superpotential,
which now includes instead of the $\mu$-term:
\beq
W = -\lambda S H_u^0 H_d^0 + \ldots
\eeq
where the ellipses may refer to a self-coupling of $S$ and/or couplings
of $S$ to other non-MSSM fields.
The corresponding supersymmetry-breaking Lagrangian is
\beq
-{\cal L}_{\rm soft} = -a_\lambda S H_u^0 H_d^0 + m_S^2 |S|^2 + \ldots
\eeq
Consider the limit of small $\lambda$, so that the resulting
theory describes a nearly unmixed singlino and MSSM neutralinos.
Then when $S$ gets its VEV, one has effectively 
\beq
\mu = \lambda \langle S \rangle; \qquad
b = a_\lambda \langle S \rangle.
\eeq
So all of the above analysis can be repeated with $b/\mu$ replaced
by $a_\lambda/\lambda$. The RG equations for the scalar cubic
couplings are given by adding a term $2 a_\lambda \lambda$ to each
of eqs.~(\ref{rgatyt})-(\ref{rgatauytau}), and replacing eq.~(\ref{rgbmu})
by
\beq
16 \pi^2
{d\over dt}(a_\lambda/\lambda) &=& 
6 g_2^2 M_2 + {6 \over 5} g_1^2 M_1 +
6 a_t y_t + 
6 a_b y_b +
2 a_\tau y_\tau + 8 a_\lambda \lambda
\eeq
where the effects of other couplings of $S$ on its anomalous dimension
are omitted. The 
last term is just a damping term and cannot change the sign of
$a_\lambda/\lambda$. In the limit of weakly coupled $S$, the
additional terms are inconsequential and the preceding analysis goes
through without change. Of course, one must still look at the details of
the particular model to decide whether it can be viable.

The above results were obtained assuming that gaugino masses are unified
to a common value $m_{1/2}$ and that scalar squared masses are unified to 
$m^2_0$. The dependence on the latter assumption is not very strong, as
the non-holomorphic scalar squared masses mainly enter into the 
determination of the sign of $\mu$ 
through their influence on $\tan\beta$, and all values of $\tan\beta$
were considered. The assumption of gaugino mass unification is stronger,
since non-unified gaugino masses will affect the running of $b/\mu$ and
$a_f/y_f$ in different ways. However, gaugino masses can be reconstructed
with good accuracy from future measurements of gluino, neutralino and 
chargino masses, so a similar analysis can be repeated for the case that
gaugino mass unification is badly violated. The top and bottom Yukawa 
couplings may well be modified from their extrapolated behavior at high
mass scales, but the dependence of Yukawa couplings on the RG evolution
comes mainly from lower scales anyway in models of gaugino mass dominance.

After the discovery of supersymmetry, it will be an important challenge
to connect measured properties of the superpartners to candidate theories
of supersymmetry breaking. In fact, there are already a couple of
weak indirect
hints from experiment which may suggest that if 
superpartners are not too
heavy and gaugino masses have a common phase, then $\mu$ should be
positive in the standard convention. First, it is often easier to 
accommodate
constraints on $b\rightarrow s \gamma$ within simple model frameworks if
$\mu > 0$ 
\cite{Baer:1998jq}. Second,
the recent measurement \cite{gminus2measurement} of the muon magnetic
dipole moment also favors this
sign \cite{gminus2signprophets,gminus2signambulancechasers} if 
$\tan\beta$ is not too small and superpartners are not too heavy.
While
caution is certainly called for before
hailing the muon $g-2$ discrepancy as evidence in favor of supersymmetry,
it should
be remembered that many models with $\mu < 0$ are
ruled out by the data at a far higher confidence level. In any
case, these
considerations highlight the importance of understanding the sign of
$\mu$ as a consequence of theory, rather than merely an input parameter.
As I have emphasized in this paper, the
theory of the mechanism of supersymmetry breaking can predict the sign of
$\mu$ in
addition
to the more obvious mass hierarchies in the sparticle spectrum.

\bigskip \noindent {\it Acknowledgements:}   This work was
supported in part by the National Science Foundation
grant number PHY-9970691. I thank Graham Kribs, Martin Schmaltz
and James Wells for helpful conversations.


\end{document}